\documentclass[twocolumn,pre,a4paper]{revtex4}
\usepackage{graphicx}
\usepackage{amsmath}
\usepackage{psfrag}
\usepackage{latexsym}

\newcommand{\bq}{\begin{equation}}
\newcommand{\eq}{\end{equation}}
\newcommand{\ba}{\begin{eqnarray}}
\newcommand{\ear}{\end{eqnarray}}

\newcommand{\X}{\mathbf{X}}
\newcommand{\x}{\mathbf{x}}
\newcommand{\ec}{\mathbf{c}}
\newcommand{\C}{\mathbf{C}}
\newcommand{\Tr}{\mbox{Tr}}
\renewcommand{\Im}{\rm{Im}\,}
\renewcommand{\Re}{\rm{Re}\,}

\begin{document}
\title{\bf Correlated Wishart Matrices and Critical Horizons}
\author{Zdzis\l{}aw Burda}
\author{Andrzej G\"orlich}
\author{Jerzy Jurkiewicz}
\author{Bart\l{}omiej Wac\l{}aw}

\affiliation{Mark Kac Center for Complex Systems Research and
Marian Smoluchowski Institute of Physics, \\
Jagellonian University, Reymonta 4, 30-059 Krak\'ow, Poland }

\begin{abstract}{ 
We discuss a practical method to determine
the eigenvalue spectrum of the empirical correlation
matrix. The method is based on the analysis of
the behavior of a conformal map
at a critical horizon which is defined as a border line 
of the physical Riemann sheet of this map. The map is a
convenient representation of the Mar\v{c}enko-Pastur 
equation.}
\end{abstract}
\maketitle

The relation between the eigenvalue spectrum of
the empirical covariance matrix and the spectrum
of the underlying covariance matrix 
plays an important role in many research areas 
ranging from physics, telecommunication \cite{m}, 
to information theory \cite{s} and quantitative finance \cite{bclp}, 
and has attracted a great attention recently.

The problem can be mathematically formulated as follows.
Suppose we have a statistical system with $N$ degrees 
of freedom $\x = (x_i)$, $i=1,\dots,N$. We collect
$T$ independent samples of $\x$,
each being a certain realization
$\X_\alpha = (X_i)_\alpha$ of $\x$, $\alpha=1,\dots,T$.
The data are stored in
a rectangular matrix $\X=(X_{i\alpha})$ of
dimension $N\times T$. The covariance
matrix\footnote{For simplicity we 
will assume that $\forall i:\;\langle x_i \rangle = 0$.}
$\C$: $C_{ij} = \langle x_i x_j \rangle$
can be estimated from the data as 
$\ec=(1/T)\X\X^{\tau}$: $c_{ij} = (1/T) \sum_t X_{it}X_{jt}$,
where $\ec$ is empirical correlation matrix.
Here $\X^\tau$ stands for the transpose of $\X$.
For $r\equiv N/T\rightarrow 0$ ($N$=const, $T\rightarrow \infty$)
the empirical covariance matrix $\ec$
perfectly approximates $\C$. However, in practice, the number
of samples $T$ is finite. One is therefore interested
in the relation between $\ec$ and $\C$,
in particular how well the eigenvalue distribution  
of $\ec$ approximates the eigenvalue distribution
of $\C$ for finite $r$.

One can formulate the problem in terms of random matrix
theory \cite{bgjj,ms}. One can think of   
$\ec=(1/T)\X\X^{\tau}$ as a matrix constructed of 
real\footnote{One can also consider
complex matrices. The large $N$-limit is in this case
identical as for real matrices.}
gaussian random matrix $\X$. 
The only requirement which one has
to impose on the probability measure for $\X$
is that the two-point correlations are:
\bq
\left\langle X_{i\alpha}X_{j\beta}\right\rangle = \delta_{\alpha\beta} C_{ij}.
\label{AC}
\eq
Such an ensemble of matrices $\X$
is called correlated Wishart ensemble \cite{w,bgjj,ms}.
The delta $\delta_{\alpha\beta}$ in the last equation tells us that 
the samples are uncorrelated, while $C_{ij}$ that the degrees
of freedom are correlated according to
the covariance matrix $\C$.

The relation between the spectral density of $\ec$ and $\C$ 
is given by the Mar\v{c}enko-Pastur equation \cite{mp}. 
This equation
has been intensively studied in the mathematical 
literature \cite{bs,cs}. The corresponding equations
for correlated samples, that
is for the case when the delta $\delta_{\alpha\beta}$ is
replaced by a symmetric matrix $A_{\alpha\beta}$ in (\ref{AC}),
have been derived recently \cite{bjw} 
using a diagrammatic technique \cite{fz,ms}.

The purpose of the present paper is twofold.
Firstly, we want to describe a practical method to calculate the eigenvalue 
spectrum of the estimator $\ec$.
Some similar methods have been discussed in the
literature \cite{bs,cs}. However, we believe that
a conformal map representation \cite{bgjj} used here 
leads to a particularly simple and effective practical method.
Secondly, the eigenvalue smearing and noise dressing 
observed in the empirical correlation matrix $\ec$
have a simple interpretation in terms of the conformal map,
which we want to present.

In order to determine the relation between the
eigenvalue densities of the covariance matrices
$\ec$ and $\C$: $\rho_{\ec}(x)$ and $\rho_{\C}(x)$
it is convenient to introduce Green's functions (resolvents):
$G(z) = (1/N) \Tr \; (z-\C)^{-1}$, 
where $z$ is a complex variable, and correspondingly 
$g(z) = (1/N) \langle \Tr \; (z-\ec)^{-1} \rangle$,
where the average is over $\X$ from a Wishart ensemble with 
the correlations given by (\ref{AC}).
It is also convenient to introduce generating functions for
the spectral moments \cite{bgjj}:
\bq
M(z) = \sum_{k=1}^\infty \frac{M_{\C k}}{z^k}, \;\; 
m(z) = \sum_{k=1}^\infty \frac{m_{\ec k}}{z^k} ,
\label{MCmc}
\eq
where $M_{\C k} = \int \rho_\C (\lambda) \lambda^k d\lambda$ 
and analogously for $m_{\ec k}$.
The generating functions are directly related to the 
resolvents: $M(z) = z G(z) - 1$ and $m(z) = z g(z) - 1$.
  
One can show \cite{bgjj} that if $N,T\rightarrow\infty$ with 
fixed $r=N/T$, the function $m(z)$ can be calculated
from $M(Z)$ as follows:
\bq
m(z) = M(Z), 
\label{mM}
\eq
where:
\bq
z=Z(1+r M (Z)).
\label{map}
\eq
Assume that we know the correlation matrix $\C$.
We can calculate eigenvalues of $\C$,
$\lambda_k$ and their multiplicities $n_k$ (degeneracies),
and determine $M(Z)$:
\bq
M(Z) = \sum_{k=1}^K
\frac{p_k \lambda_k}{Z-\lambda_k},
\label{MCZ}
\eq
where $p_k = n_k/N$ is the fraction of all
eigenvalues taking the value $\lambda_k$.
For the later convenience we also assume that
the eigenvalues are ordered $\lambda_1<\dots < \lambda_K$.

For the given $M(Z)$ we can apply the equations (\ref{mM}) and (\ref{map}) 
to determine $m(z)$ and $g(z) = (1+m(z))/z$ and further, from $g(z)$
we can calculate the eigenvalue 
density $\rho_{\ec}(x)$ by taking the imaginary part of $g(z)$
along the cuts on the real axis: 
$\rho_\ec(x) = -(1/\pi) \Im\,g(x+i0^+)$. 

The equations (\ref{mM}) and (\ref{map}) are equivalent 
to the Mar\v{c}enko-Pastur equation \cite{mp}. 
Let us now discuss their physical content.

\onecolumngrid

\begin{figure}
\psfrag{Imz}{$\Im z$} \psfrag{Rez}{$\Re z$}
\psfrag{ImZ}{$\Im Z$} \psfrag{ReZ}{$\Re Z$}
\psfrag{Z}{$Z$} \psfrag{Zp}{$Z'$}
\psfrag{z=z(Z)}{$z=z(Z)$} \psfrag{Z=Z(z)}{$Z=Z(z)$} \psfrag{z=zp}{$z=z'$}
\psfrag{s}{$\sqrt{r}$}
\psfrag{1}{$1$} \psfrag{1+r}{$1+r$} \psfrag{l-}{$x_-$} \psfrag{l+}{$x_+$}

\includegraphics[width=15cm]{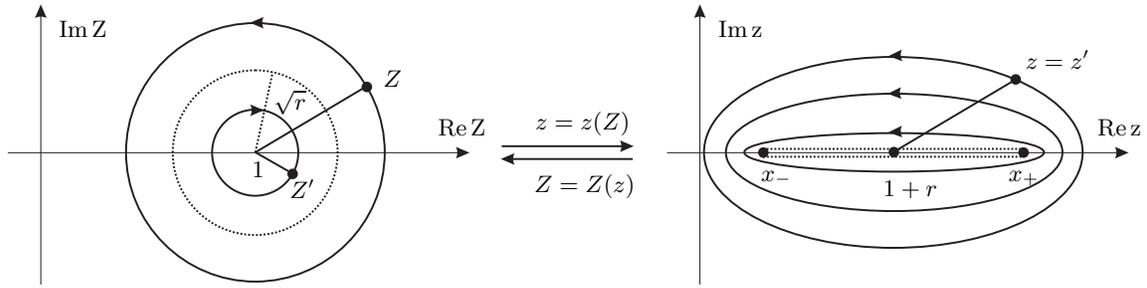}
\caption{Graphical representation of the map $z=z(Z)$. 
Each pair of points: $Z=1+R e^{i\phi}$ and $Z'=1+(r/R) e^{-i\phi}$ 
is mapped onto the same point $z=z'$.
While $\phi$ increases, $Z$ and $Z'$ 
move along the circles in the directions given by arrows.
The critical horizon (dotted circle) has radius $\sqrt{r}$.
The image of a circle $|Z-1| = R$ is an ellipse 
in $z$-plane, which for $R\rightarrow \sqrt{r}$ 
degenerates to an interval (dotted line).}
\label{fig1a}
\end{figure}

\begin{figure}
\psfrag{Imz}{$\Im z$} \psfrag{Rez}{$\Re z$}
\psfrag{ImZ}{$\Im Z$} \psfrag{ReZ}{$\Re Z$}
\psfrag{z=z(Z)}{$z=z(Z)$} \psfrag{Z=Z(z)}{$Z=Z(z)$}
\includegraphics[width=15cm]{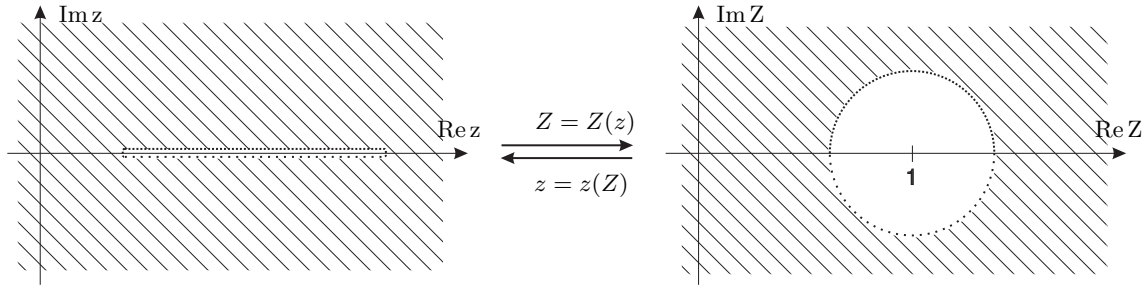}
\caption{The inverse function $Z=Z(z)$: the $z$-plane 
without the real interval $[x_-,x_+]$ is mapped onto 
the outside of the critical circle in the $Z$-plane.
The interval is mapped into the upper (or lower)
part of the critical semicircle.}
\label{fig1b}
\end{figure}

\twocolumngrid

The equation (\ref{mM}) tells us that 
the function $m(z)$ assumes for $r=0$ the
same form as the function $M(Z)$ since in this case
$z=Z$ as follows from the Eq. (\ref{map}).

In this case the spectrum of the experimental
matrix $\ec$ is identical as of the matrix $\C$,
as expected for $T\rightarrow\infty$. However for $r>0$, 
$z=z(Z)$ is a nontrivial function and the relation
of the generating function $m(z)$ to the original one
$M(Z)$ is not simple. The function $M(Z)$ has poles 
on the real axis at $\lambda_k$'s. We may ask 
whether one can see the traces of the poles
also in the function $m(z)$.

In order to answer this question 
first consider the simplest example 
$\lambda_1=\lambda_2=\ldots=\lambda_N=1$. The function $M(Z) = 1/(Z-1)$ 
has only one pole and the
conformal map (\ref{map}) takes the form:
\bq
z = Z\left(1 + \frac{r}{Z-1}\right).
\label{zZ1}
\eq
Consider a point on the $Z$-plane in the distance $R$ from the
pole: $Z=1+R e^{i\phi}$. The function 
(\ref{zZ1}) assumes the value $z=1 + r + Re^{i\phi} + (r/R) e^{-i\phi}$ 
the same as for a point $Z' = 1 + (r/R) e^{-i\phi}$.
As a consequence each point $Z$ outside 
the circle $|Z-1| = \sqrt{r}$ has a partner $Z'$ 
inside this circle such that $z(Z)=z(Z')$ as depicted in Fig. \ref{fig1a}.
On the limiting circle there are pairs of points: $Z=1+\sqrt{r} e^{i\phi}$
and $Z'=\bar{Z}=1+\sqrt{r} e^{-i\phi}$
for which $z$ assumes the same values.
We see that the upper part of the 
semicircle is mapped onto an interval
$[x_-,x_+],\; x_\pm=(1\pm\sqrt{r})^2$ on the real axis in the $z$-plane,
and so is the lower semicircle. 
Inverting the function $z=z(Z)$ one obtains a two-valued
function $Z=Z(z)$ and thus
one has to decide which Riemann sheet to choose.
The physical Riemann sheet corresponds to the outside of the circle
$|Z-1|=\sqrt{r}$ as follows from the expansion (\ref{MCmc}).
For the inverse function $Z=Z(z)$ the image of the whole $z$-plane 
is given by the outside of the circle $|Z-1| = \sqrt{r}$
plus the upper semicircle which is an image of the cut (see Fig. \ref{fig1b}). 
We thus see that the pole in the $Z$-plane is surrounded 
by a critical horizon behind which the argument of the inverse function
$Z=Z(z)$ never enters. Only when the horizon radius $\sqrt{r}$ 
shrinks to zero the variable $Z$ can approach the ``naked'' pole.

As mentioned before, both semicircles of the
critical horizon $|Z-1| = \sqrt{r}$ are mapped by 
the transformation (\ref{map}) onto the same interval
$[x_-,x_+]$ on the real axis which makes 
the inverse function $Z=Z(z)$ two-valued. 
To make out of it a single-valued one one has to 
restrict the image to either the upper or lower semicircle.
This can be done by approaching the cut
from above: $z=x+i0^+$ to 
obtain the upper semicircle or from below $z=x+i0^-$
for the lower semicircle (Fig. \ref{fig1b}).
The imaginary part of $g(z)=\left(1+M(Z(z))\right)/z$ 
gives, for $z=x+i0^+$, the eigenvalue density $\rho_{\ec}(x)$.

\begin{figure}
\psfrag{xx}{$X$} \psfrag{yy}{$Y$}
\psfrag{rho}{$\rho_\ec(x)$} \psfrag{lambda}{$x$}
\includegraphics[width=8cm]{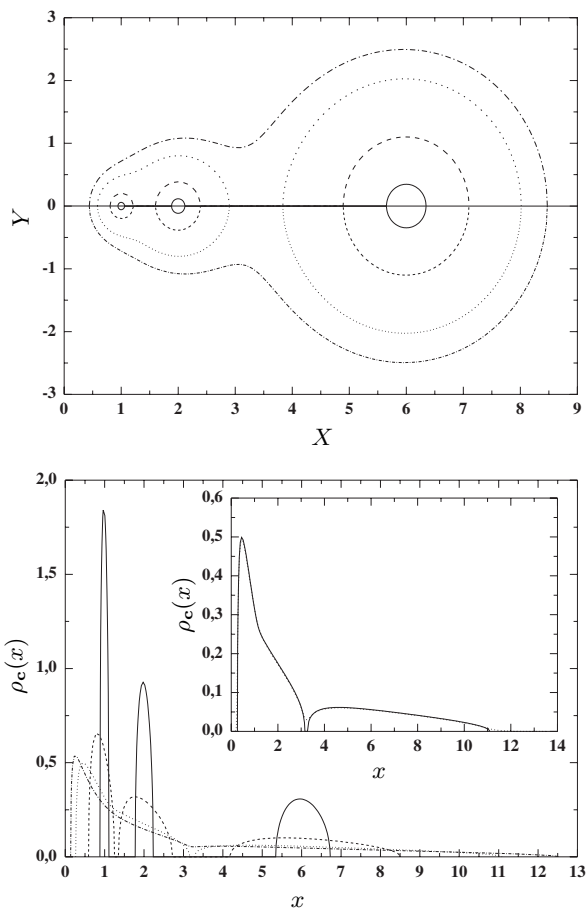}
\caption{Top: Critical horizon 
for $\{\lambda_i\}=\{1,2,6\}$ and for 
$r=1/100$ (solid), $r=1/10$ (dashed),
$r=1/3$ (dotted) and $r=1/2$ (dash-dot). 
Bottom: The corresponding spectra $\rho_\ec(x)$. Inset: the spectrum for $r=1/3$, calculated (solid line)
and found experimentally (dotted) for sample of $3\times 10^5$ matrices of size $N=48$.}
\label{fig2}
\end{figure}

We see that the shape of the eigenvalue density $\rho_{\ec}(x)$
is encoded in the behavior of the conformal
map (\ref{map}) near the critical horizon.
This is true in the general case (\ref{MCZ}). The poles
of $M(Z)$ are distributed on the positive real semi-axis 
at $\lambda_k$'s. For $r>0$ they are shielded behind 
a critical horizon (see Fig. \ref{fig2}). 
In general there are $K+1$ Riemann sheets.
We are interested in the one for which
$1/z\rightarrow 0$ when $1/Z \rightarrow 0$ in Eq.
(\ref{MCmc}). Thus
the physical region lies outside the critical horizon.
It is symmetric about the real axis.
The upper part ($\Im Z>0$) is mapped by Eq. (\ref{map})
onto intervals on the real axis in the $z$-plane and so is
the lower one. When $r$ is positive but very
small each pole has its own horizon, but when $r$ is increased
the individual horizons grow and merge so that the number of
connected components of the total horizon decreases with $r$ 
(see Fig. \ref{fig2}).
For sufficiently large $r$ a single critical horizon surrounds all poles.
The corresponding image of the horizon 
on the $z$-plane has a single cut on the real axis.
For the limiting case $r=1$, the cut touches the origin at $z=0$.

The discussion presented above can be turned into a practical
method of determining the shape of the eigenvalue distribution
$\rho_\ec(x)$. We have to find the
inverse function $Z=Z(z)$ to $z=z(Z)$ (\ref{map}), insert the solution
to $M(Z)$ and determine the behavior of the resulting function
along the cuts on the real axis. The equation (\ref{map}) can
be analytically solved for $Z$ only in few cases.
In the general case one has to use a numerical method.
Fortunately, one can bypass the problem of the explicit 
function's inversion by using a method described below. 

In the first step we determine
the critical horizon. Then moving gradually along its upper part 
($\Im Z\ge 0$) we calculate $z=z(Z)$ (\ref{map}) which is in
this case real ($z=x+i0^+$)
and simultaneously 
$\hat{\rho}_\ec(Z)= -(1/\pi) \Im (M(Z)+1)/z(Z)\equiv 
-(1/\pi) \Im (m(z)+1)/z$ 
which gives the eigenvalue density
$\rho_\ec(x)=\hat{\rho}_{\ec}(Z)$.
Briefly speaking, the method is to use the auxiliary variable 
$Z$ on the upper part of the horizon
to parametrize both the eigenvalue $x=z(Z)$ and 
$\rho_\ec(Z)$, and to treat the pairs $(x(Z),\hat{\rho}_\ec(Z))$ 
as $\rho_{\ec}(x)$.
If the horizon has many disconnected parts,
also the cut and the support of the eigenvalue
density function $\rho_{\ec}(x)$
consists of many disconnected intervals. 

Let us describe how to determine the critical horizon.
If $Z=X+iY$ is a point on the horizon,
the imaginary part of the map $z=z(Z)$ (\ref{map}) 
vanishes and we have:
\bq
\sum_{k=1}^K \frac{p_k\lambda_k^2}{(X-\lambda_k)^2+Y^2} = \frac{1}{r} .
\label{main}
\eq
This equation has to be solved for $Y=Y(X)$ for given $X$.
If the real solution exists it has two symmetric roots $\pm Y$.
We are interested in the non-negative one $Y\ge 0$ which
corresponds to $Z$ on the upper part of the 
critical horizon. Now we can calculate
$x=z(Z)$ and $\rho_\ec(Z)$:
\bq
x(Z) = X + r \sum_{k=1}^K p_k \lambda_k + 
r \sum_{k=1}^K \frac{p_k\lambda_k^2(X-\lambda_k)}{
(X-\lambda_k)^2 + Y^2},
\eq
and
\bq
\rho_\ec(Z) = -\frac{Y}{\pi x} 
\sum_{k=1}^K \frac{p_k\lambda_k}{(X-\lambda_k)^2+Y^2} ,
\label{eq:rhoc3}
\eq
which, after ignoring $Z$ being a parameter,
gives the pair $(x,\rho_\ec(x))$. The two equations above
are explicit, so one can directly compute $x$ and $\rho_\ec$
for any given $Z$ if one knows the dependence $Y(X)$.
This dependence can be obtained by solving 
the Eq. (\ref{main}) for $Y(X)$. 

The solution of Eq. (\ref{main}) depends on $r$.
The function on the left-hand side of (\ref{main}) 
has a shape of a 'hilly' landscape when plotted on $Z$-plane, with $K$
peaks of infinite height located at the poles $(\lambda_k,0)$. The 
circumference of each peak shrinks to zero for increasing altitude.
The solution of the equation lies on curves obtained as
a cross-section of the hilly landscape at height $1/r$. 
One can think of a coast line around islands surrounded 
by a sea with the water level $1/r$. 
For $r\rightarrow 0^+$, the level goes to infinity, 
and thus the cross-section contains only
the points at which the peaks are located: $(\lambda_k,0)$.
For $r>0$, the level $1/r$ is finite, so the
cross-section contains closed lines surrounding the poles
and forming the critical horizon. When the water level
drops the islands merge and the coast line grows (Fig. \ref{fig2}).

One can find the points where the horizon crosses the real 
axis by setting $Y=0$ in Eq. (\ref{main}). Among those points,
the leftmost $X_-$ and the rightmost $X_+$ set the limits
on the minimal and maximal value of $X$ where one has to
look for the horizon solution.
The two points are  mapped by Eq. (\ref{map}) 
onto $x_\pm$ being the lower and upper edge of the eigenvalue 
spectrum $\rho_{\ec}(x)$. The values $X_-$ and $X_+$
can be easily found by a root finder algorithm applied
to the Eq. (\ref{main}) with $Y=0$. Having determined
$X_-$ and $X_+$ one can change $X$ 
in small steps from $X_-$ to $X_+$
and for each $X$ find the positive solution $Y$ 
of Eq. (\ref{main}) to eventually determine $Z=X+iY(X)$ 
on the horizon.  In that way the problem is solved.

Let us now discuss examples. 
First we consider a correlation matrix 
$\C$ having three eigenvalues 
$\{1,2,6\}$ with equal weights $p_k=1/3$.
The critical horizon for different $r$ as well as corresponding eigenvalue 
spectra $\rho_\ec(x)$ are shown in Fig. \ref{fig2}
The evolution of the critical horizon with $1/r$ can be
treated as a level map for the landscape (\ref{main}).
In the inset of Fig. \ref{fig2} we compare
the spectrum for $r=1/3$ with a spectrum obtained  
by diagonalization of $n=3\cdot 10^5$ matrices of size $N=48$ and $T=144$
chosen randomly from the corresponding Wishart ensemble.
The agreement is perfect, up to a finite-size corrections 
near the edges of the cut.

\begin{figure}
\psfrag{xx}{$x$} \psfrag{yy}{$\rho_\ec(x)$}
\includegraphics[width=8cm]{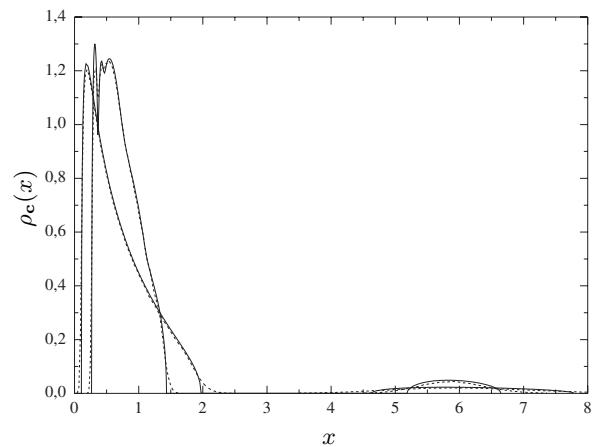}
\caption{Solid line: $\rho_\ec(x)$ for $N=18$ eigenvalues of $\C$ taken from
Polish Stock Market, for $T=54\leftrightarrow r=0.333$ and 
$T=255\leftrightarrow r=0.0706$, calculated using the method 
described in this paper. Dashed line: $\rho_{\ec}(x)$ obtained
by diagonalization of $3 \cdot 10^5$ Wishart matrices generated
by a Monte-Carlo procedure.
\label{fig3}}
\end{figure}

The second example is related to some practical problem.
The portfolio selection is
a central problem of quantitative finance.
The importance of the random matrix theory for this problem
has been recently discovered \cite{bclp}. It has been realized
that the lower part of the spectrum $\rho_\ec$ has a universal shape
stemming from statistical fluctuations.
The method which we described above 
gives a refined tool allowing one to 
observe a fine structure of the spectrum of the empirical
matrix also in its lower part. As an example
we consider a correlation matrix obtained
for returns of $18$ stocks on the Polish Stock Market.
The eigenvalues read: $0.36$, $0.38$, $0.47$, $0.48$, $0.56$, $0.59$, $0.64$, 
$0.66$, $0.68$, $0.71$, $0.78$, $0.83$,
$0.89$, $0.94$, $0.95$, $1.08$, $1.16$, $5.84$, 
they are normalized to $\sum_k \lambda_k = N$. 
In Fig. \ref{fig3} we show
the expected eigenvalue spectrum of $\ec$
for $N=18$ and for $T=54$ ($r=0.333$) and $T=255$ ($r=0.0706$).
For small $r$ a fine structure emerges in the spectrum.
In figure the spectrum is also compared to the one obtained by
diagonalization of $n=3\times 10^5$ correlated Wishart matrices. 
The agreement is very good. This means that 
already for $N$ of order $20$
the large $N$ limit, in which the analytical equations
have been derived, applies. Deviations 
are observed only at the edges of the individual parts of the
spectrum.


Statistical properties of random matrices and complex
analysis are very intertwined. We used here this
interrelation, or more precisely a conformal map representation
of the Mar\v{c}enko-Pastur equations, to derive a general,
practical method for the determination of the eigenvalue 
distribution for the standard estimator of 
the covariance matrix. 
This method can also be applied to the
corresponding equations \cite{bjw}
for the case of a correlated sample where the delta
function $\delta_{\alpha\beta}$ in Eq. (\ref{AC}) is substituted by
a correlation matrix $A_{\alpha\beta} \ne \delta_{\alpha\beta}$.

This work was partially supported by
the Polish State Committee for Scientific Research (KBN) grant
2P03B-08225 (2003-2006) and Marie Curie Host Fellowship
HPMD-CT-2001-00108 and by EU IST Center of Excellence ``COPIRA''.

\end{document}